\documentclass[prd,twocolumn,superscriptaddress,showpacs,preprintnumbers,nobibnotes,nofootinbib,]{revtex4}
\usepackage{amsmath,graphicx,epsfig}

\def\a{\alpha}

\def\g{\gamma}

\def\D{\Delta}
\def\G{\Gamma}
\def\La{\Lambda}
\def\pl{\partial}
\def\hs{\hspace}
\def\ol{\overline}
\def\no{\nonumber}
\def\lf{\left}
\def\rg{\right}
\def\be{\begin{equation}}
\def\ee{\end{equation}}

\begin{document}

\preprint{ADP-05-08/T618, JLAB-THY-05-322}

\title{Nucleon quark distributions in a covariant quark-diquark model}

\author{I.C.~Clo\"et}
\email{icloet@physics.adelaide.edu.au}
\affiliation{Special Research Centre for the Subatomic Structure of Matter and \\
             Department of Physics and Mathematical Physics, University of Adelaide,
             SA 5005, Australia}
\affiliation{Jefferson Lab, 12000 Jefferson Avenue, Newport News, VA 23606, U.S.A.}
\author{W.~Bentz}
\email{bentz@keyaki.cc.u-tokai.ac.jp}
\affiliation{Department of Physics, School of Science, Tokai University, 
                         Hiratsuka-shi, Kanagawa 259-1292, Japan}
\author{A.W.~Thomas}
\email{awthomas@jlab.org}
\affiliation{Jefferson Lab, 12000 Jefferson Avenue, Newport News, VA 23606, U.S.A.}

\begin{abstract}
Spin-dependent and spin-independent quark light-cone momentum distributions and structure functions 
are calculated for the nucleon. We utilize a modified Nambu$-$Jona-Lasinio model
in which confinement is simulated by eliminating unphysical thresholds for nucleon
decay into quarks. The nucleon bound state
is obtained by solving the Faddeev equation in the quark-diquark approximation, where both 
scalar and axial-vector diquark channels are included. We find excellent agreement 
between our model results and empirical data.
\end{abstract}

\pacs{13.60.Hb,12.38.Lg,11.80.Jy,12.39.Fe,12.39.Ki}
\maketitle

\section{Introduction}

The discovery in the late 1980's by the European Muon Collaboration (EMC)
that the fraction of the spin of the proton carried by the quarks 
is unexpectedly small \cite{Ashman:1987hv}, caused much excitement 
in the nuclear and particle physics communities.  
The ``proton spin crisis'' prompted many new experiments, leading to major
new insights into the spin structure of the proton. 
However, a thorough theoretical understanding of the non-perturbative parton 
distributions still remains an exciting challenge.

In this paper we calculate the spin-independent and spin-dependent
quark distributions in the framework developed by Bentz, Thomas
and collaborators, in which proper-time regularization is applied to
the Nambu$-$Jona-Lasinio (NJL) model \cite{Nambu:1961tp} in order to simulate the 
effects of confinement~\cite{Mineo:2003vc}.
This model is attractive because of its covariance and
the transparent description of spontaneous chiral symmetry breaking.
We construct the nucleon as a bound state solution of
the relativistic Faddeev equation~\cite{Huang:1993yd,Ishii:1995bu,Oettel:2000jj,Maris:2003vk}  in the quark-diquark
approximation \cite{Bentz:2001vc},
where both scalar and axial-vector diquark channels are included.
This quark-diquark description of the single nucleon has the
further advantage that it can be extended to finite 
baryon density~\cite{Cloet:2005rt}.
We pay special attention to the polarized structure of the nucleon,
comparing our results for the quark distributions
with the empirical parameterizations.

\section{Quark distributions}

The spin-dependent quark light-cone momentum distribution in the nucleon,
at leading twist, is defined by Eq.~(\ref{eqn:defQD}),  
where $\psi_q$ is the quark field of flavour $q$, $x$
is the Bjorken scaling variable
and the subscript $c$ reminds us that only connected
matrix elements are included.
\begin{multline}
\D f_q(x) = p_- \, \int \frac{d\xi^-}{2\,\pi}\,e^{i\,x\,p^+\,\xi^-} \\
\langle p,s \lvert \ol{\psi}_q(0)\,\g^+\g_5 \psi_q(\xi^-)\rvert p,s\rangle_c.
\label{eqn:defQD}
\end{multline}
We normalize the nucleon state vector according to 
non-covariant light-cone normalization: 
$\langle p,s \lvert \ol{\psi}_q \,\g^+ \psi_q \rvert p,s\rangle_c = 3$. 
The spin-independent
distribution, $f_q(x)$, is defined by the 
replacement $\gamma^+ \gamma_5 \to \gamma^+$ in Eq.~(\ref{eqn:defQD}).
To determine the quark distributions in this model,
it is convenient to express Eq.~(\ref{eqn:defQD}) 
in the form \cite{Jaffe:1985je,Barone:2001sp}
\begin{multline}
\D f_q(x) = -i\int \frac{d^4 k}{\lf(2\pi\rg)^4} \\
\delta\!\lf(x - \frac{k_-}{p_-}\rg) \textrm{Tr}\biggl(\g^+\g_5\,M_q\lf(p,k\rg)\biggr),
\label{eqn:def2}
\end{multline}
where $M_q\lf(p,k\rg)$ is the quark two-point function in the bound nucleon. 
Hence, within any model that describes the nucleon as a bound state of quarks,
the distribution functions can be associated with a straightforward Feynman
diagram calculation. 

\begin{figure}[tbp]
\centering\includegraphics[width=\columnwidth,angle=0]{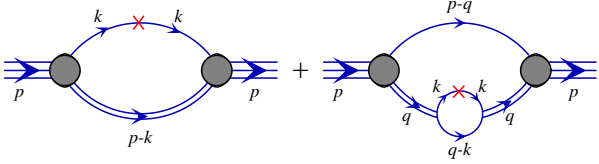}
\caption{Feynman diagrams representing the quark distributions in the nucleon, 
needed in the evaluation of Eq.~(\ref{eqn:defQD}). The single line
represents the quark propagator and the double line the 
diquark $t$-matrix. The shaded oval
denotes the quark-diquark vertex function and the operator insertion has the form
$\gamma^+\gamma_5\,\delta\!\!\left(x - \frac{k_-}{p_-}\right)\frac{1}{2}\left(1 \pm \tau_z\right)$
for the spin-dependent distribution and $\g^+\g_5 \to \g^+$ for the spin-independent case.}
\label{fig:feydiagrams}
\end{figure}

The Feynman diagrams considered here are given in Fig.~\ref{fig:feydiagrams},
where in our model the resulting distributions have no support for negative $x$.
Therefore this is essentially a valence quark picture.
By separating the isospin factors, the spin-dependent
$u$ and $d$ distributions in the proton can be expressed as 
\begin{align}
\label{eqn:delu}
\D u_v(x) &= \D f^s_{q/N}(x) + \frac{1}{2}\,\D f^s_{q(D)/N}(x) + \frac{1}{3}\, \D f^a_{q/N}(x)\no \\
&\hs{6mm} + \frac{5}{6}\,\D f^a_{q(D)/N}(x) + \frac{1}{2\sqrt{3}} \D f_{q(D)/N}^m (x),  \\ 
\label{eqn:deld}
\D d_v(x) &= \frac{1}{2}\,\D f^s_{q(D)/N}(x) + \frac{2}{3}\,\D f^a_{q/N}(x)  \no \\
&\hs{6mm} + \frac{1}{6}\,\D f^a_{q(D)/N}(x) - \frac{1}{2\sqrt{3}} \D f_{q(D)/N}^m (x).
\end{align}
The superscripts $s$, $a$ and $m$ refer to the scalar, axial-vector or mixing
terms, respectively, the subscript $q/N$ implies a quark diagram and similarly
$q(D)/N$ a diquark diagram. 
Because the scalar diquark has spin zero, 
we have $\D f^s_{q(D)/N}(x)=0$ and hence the
polarization of the $d$-quark arises exclusively from 
the axial-vector and the mixing terms. 
Similar expressions hold for the 
spin-independent distributions, but in that case
there is no mixing contribution ($f_{q(D)/N}^m=0$) \cite{Mineo:2002bg}. 

Importantly, in this covariant framework, the Ward identities corresponding to
number and momentum conservation are satisfied, 
guaranteeing the validity of the  
baryon number and momentum sum rules \cite{Mineo:1999eq,Mineo:2003vc}.

\section{The nucleon in the NJL model}

The NJL model is a chiral effective quark theory 
that is characterized by a 4-Fermi
contact interaction of the form, 
${\cal L}_I = \sum_i G_i \,\lf(\ol{\psi}\,\G_i\,\psi\rg)^2$,
where the $\G_i$ represent matrices in Dirac, 
colour and flavour space and $G_i$ are
coupling constants~\cite{Nambu:1961tp}. Applying Fierz transformations, 
the interaction Lagrangian can be decomposed into various
interacting $q \bar{q}$ and $qq$ channels.
Writing only those terms relevant to this discussion, we have
\begin{equation}
{\cal L} = \ol{\psi}\lf(i\!\! \not\!\pl - m\rg)\psi + {\cal L}_{I,\pi} + {\cal L}_{I,s} + {\cal L}_{I,a},
\end{equation}
where $m$ is the current quark mass. The interaction terms are given by
\begin{align}
&\hs{-1.5mm}{\cal L}_{I,\pi} = \frac{1}{2}\, G_\pi \lf(\lf(\ol{\psi}\psi\rg)^2 - 
\lf(\ol{\psi}\,\g_5\vec{\tau}\,\psi \rg)^2\rg), \\
&\hs{-1.5mm}{\cal L}_{I,s}   = G_s \Bigl(\ol{\psi}\,\g_5 C \tau_2 \beta^A\, \ol{\psi}^T\Bigr)
\Bigl(\psi^T\,C^{-1}\g_5 \tau_2 \beta^A\, \psi\Bigr),\\
&\hs{-1.5mm}{\cal L}_{I,a}   = G_a \Bigl(\ol{\psi}\,\g_\mu C \tau_i\tau_2 \beta^A\, \ol{\psi}^T\Bigr)
\Bigl(\psi^T\,C^{-1}\g^{\mu} \tau_2\tau_i \beta^A\, \psi\Bigr),
\label{eqn:Lint}
\end{align}
where $\beta^A = \sqrt{\frac{3}{2}}\,\lambda^A~(A=2,5,7)$
are the colour $\ol{3}$ matrices and $C = i\g_2\g_0$.
The familiar term ${\cal L}_{I,\pi}$ generates the constituent quark mass, $M$, via
the gap equation and the pion as a $q \bar{q}$ bound state.   
The terms ${\cal L}_{I,s}$ and ${\cal L}_{I,a}$ 
represent the interactions in the
scalar ($J^{\pi} = 0^+, T = 0, \text{colour}\,\ol{3}$) and axial-vector 
($J^{\pi} = 1^+, T = 1, \text{colour}\,\ol{3}$) diquark 
channels and are used to construct the
nucleon as a quark-diquark bound state. 
The couplings $G_\pi$, $G_s$ and $G_a$ are related to 
the original couplings, $G_i$,
via the Fierz transformation, but we use them here as free 
parameters which will be fixed by the properties
of the pion and the nucleon. 
 
Solving the appropriate Bethe-Salpeter equations,  
the standard NJL results for the diquark $t$-matrices are 
obtained \cite{Ishii:1995bu,Mineo:2002bg}.  
As explained in Ref.~\cite{Bentz:2001vc}, these can be accurately approximated
by the forms
\begin{align}
\label{taus}
\tau_{s}(q) &=  4i \, G_s\, - \frac{i g_s} {q^2 - M_s^2}, \\
\tau_a^{\mu\nu}(q) &= 4i G_a\, g^{\mu\nu} - \frac{i g_a}{q^2 - M_a^2} 
\left(g^{\mu\nu} - \frac{q^{\mu}q^{\nu}}{M_a^2} \right),
\label{taua}
\end{align}
which we also use here.
The masses of the diquarks $M_s, \, M_a$ and their
couplings to the quarks $g_s, \, g_a$ are defined as the poles and residues of 
the appropriate full diquark $t$-matrices. 

The nucleon (quark-diquark) $t$-matrix satisfies the Faddeev equation 
\be
T =  Z + Z\,\Pi_N\,T =  Z + T\,\Pi_N\,Z,
\label{eqn:t}
\ee 
where $Z$ is the quark exchange kernel and $\Pi_N$ the product of a quark propagator
and a diquark $t$-matrix. 
In the non-covariant light-cone normalization used already
in Eq.~(\ref{eqn:defQD}), the quark-diquark vertex function, $\Gamma_N$, is defined by the 
behaviour of $T$ near the pole
\begin{align} 
T \stackrel{p_+ \to \varepsilon_p}{\longrightarrow} 
\frac{\Gamma_N\,\overline{\Gamma}_N}{p_+ - \varepsilon_p},
\label{pole}
\end{align}
where $\varepsilon_p = \frac{M_N^2}{2p_-}$ is the light-cone energy. Substituting
this result into Eq.~(\ref{eqn:t}) gives the homogeneous Faddeev equations for
the vertex functions  
\begin{align}
\label{eqn:f}
\G_N = Z\,\Pi_N \, \G_N, \quad \text{and} \quad
\ol{\G}_N = \ol{\G}_N\, \Pi_N \,Z.
\end{align}

For this investigation we restrict ourselves to the static approximation, where 
we neglect the momentum dependence of the quark exchange kernel, $Z$.
Including both scalar and axial-vector diquark channels, $Z$ takes the following 
form in the colour singlet and isospin-$\tfrac{1}{2}$ channel:
\begin{align}
Z = \frac{3}{M} \begin{pmatrix} 1 & \sqrt{3}\g_{\mu'}\g_5 \\
             \sqrt{3}\g_5\g^{\mu} & -\g_{\mu'}\g^{\mu} \end{pmatrix}.
\end{align}
The quantity $\Pi_N$ effectively becomes the quark-diquark bubble graph:
\begin{align}
\Pi_N(p) &= \int \frac{d^4k}{(2\pi)^4}\, \tau(p-k)\, S(k),
\end{align}
where
\begin{align}
\tau(q) =  \begin{pmatrix} \tau_s(q) & 0 \\ 0 & \tau_a^{\mu\nu}(q) \end{pmatrix}.
\label{tau}
\end{align}
The eigenfunction of the kernel $K \equiv Z\,\Pi_N$, in Eq.~(\ref{eqn:f}), has the following form, 
up to normalization:
\begin{align}
\G(p,s) = \begin{bmatrix} \a_1 \\ \a_2\,\frac{p^{\mu}}{M_N}\,\g_5\ + \a_3\,\g^\mu\g_5 \end{bmatrix}u_N(p,s),
\label{eqn:nvertex}
\end{align}
where the upper and lower component refer to the scalar and axial-vector diquark channels, respectively
and $u_N$ is a free Dirac spinor with mass $M_N$. We choose the
normalization $\ol{u}_N u_N = 1 = \ol{\G}\G$.\footnote{The conjugate vertex function, $\ol{\G}$, 
which is a left eigenfunction of $\ol{K} \equiv \Pi_N\,Z$, 
is obtained by taking the ordinary hermitian conjugate of $\Gamma$ and introducing a minus sign for the
axial-vector components.} Inserting this form into Eq.~(\ref{eqn:f}) gives three homogeneous equations for the
$\alpha$'s and the nucleon mass $M_N$ is determined by the requirement that the eigenvalue
of $K$, in Eq.~(\ref{eqn:f}), equal 1. 

The normalization of the vertex function follows from the definition given in Eq.~(\ref{pole}), we obtain
\begin{align}
\Gamma_N(p,s) &= \sqrt{-Z_N\frac{M_N}{p_-}}\,\G(p,s),
\end{align}
where
\begin{align}
Z_N = \frac{p_-}{M_N}\,\frac{-1}{\G(p)\,\frac{\pl \Pi_N(p)}{\pl p_+}\,\G(p)}.
\end{align}

As with any non-renormalizable theory a regularization prescription
must be specified to fully define the model. 
We choose the proper-time regularization
scheme \cite{Schwinger:1951nm,Ebert:1996vx,Hellstern:1997nv,Bentz:2001vc}, 
where loop integrals of products of propagators are evaluated by
introducing Feynman parameters, Wick rotating and making the denominator 
replacement 
\begin{equation}
\frac{1}{X^n} 
\longrightarrow 
\frac{1}{(n-1)!}\,\int_{1/(\La_{UV})^2}^{1/(\La_{IR})^2}\,d\tau\,\tau^{n-1}\,e^{-\tau\,X},
\end{equation}
where $\La_{IR}$ and $\La_{UV}$ are, respectively, 
ultraviolet and infrared cutoffs.
The former has the effect of
eliminating unphysical thresholds for hadron decay into
quarks, hence simulating an important aspect of confinement \cite{Hellstern:1997nv}.

\section{Results}

The parameters of the model are $\La_{IR}$, $\La_{UV}$, $m$, $G_\pi$, 
$G_s$ and $G_a$. The infrared scale is expected to be of order $\La_{QCD}$ 
and we set it to $\La_{IR} = 0.28\,$GeV. This is slightly larger than 
our previous work \cite{Mineo:2003vc}, because our studies of the saturation 
properties of nuclear matter favour this~\cite{Cloet:2005rt}.
The parameters $m$, $\La_{UV}$ and $G_\pi$ are determined by requiring 
$M=400\,$MeV via the gap equation, $f_\pi=93\,$MeV from the familiar
one loop pion decay diagram and $m_\pi=140\,$MeV from 
the pole of the $q \overline{q}$ $t$-matrix
in the pion channel. This gives $m=15.3\,$MeV, 
$\La_{UV} = 0.66\,$GeV and $G_\pi = 17.81\,$GeV$^{-2}$.
The couplings $G_s$ and $G_a$ are determined by reproducing the nucleon mass
$M_N = 940\,$MeV as the solution of Eq.~(\ref{eqn:f}) and 
satisfying the Bjorken sum rule within our model, where $g_A = 1.267$. 
We obtain $G_s = 8.41~$GeV$^{-2}$ and $G_a = 1.36~$GeV$^{-2}$. 
With these model parameters the diquark masses 
are $M_s= 0.65\,$GeV and $M_a=1.2\,$GeV
and the coefficients in the
nucleon vertex function, Eq.(\ref{eqn:nvertex}), 
are $\lf(\a_1,\a_2,\a_3\rg) = \lf(-0.35,-0.0088,0.47\rg)$.

To compare the predictions of the model with experimental data as well
as the empirical parameterizations, it is necessary to determine the 
model scale, $Q_0^2$. We do this by 
optimizing $Q_0^2$ such that the spin-independent distribution, $u_v(x)$, 
best reproduces the empirical parameterization after $Q^2$ evolution.
We find a model scale of $Q_0^2 = 0.16~$GeV$^2$, which is typical of
valence dominated models \cite{Mineo:2002bg,Mineo:1999eq,Schreiber:1991tc}.

\begin{figure}[tbp]
\centering\includegraphics[width=\columnwidth,angle=0]{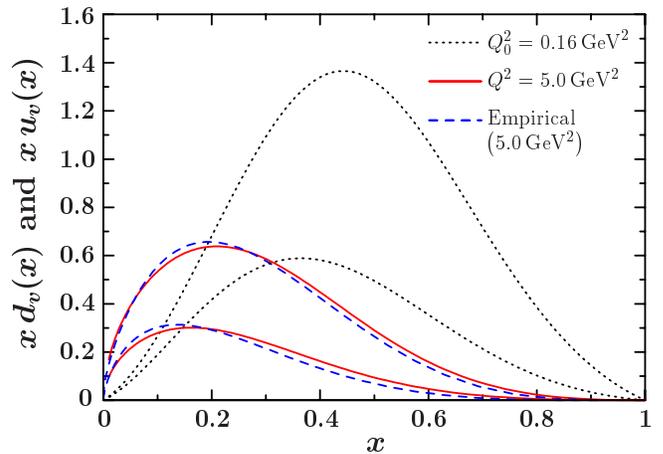}
\caption{Spin-independent valence $u$ and $d$ distributions multiplied by Bjorken $x$.
There are three curves for each quark flavour,
with the lower curve of each type representing the $d$ distribution.
The dotted line is the model prediction at
the NJL scale of $Q_0^2 = 0.16~\text{GeV}^2$ and the solid 
line is the result after QCD evolution
to the scale $Q^2 = 5.0~\text{GeV}^2$. The dashed line 
is the empirical parametrization
of Ref.~\cite{Martin:2002dr}, at the scale $Q^2 = 5.0~\text{GeV}^2$.}
\label{fig:SI_xup}
\end{figure}

\begin{figure}[tbp]
\centering\includegraphics[width=\columnwidth,angle=0]{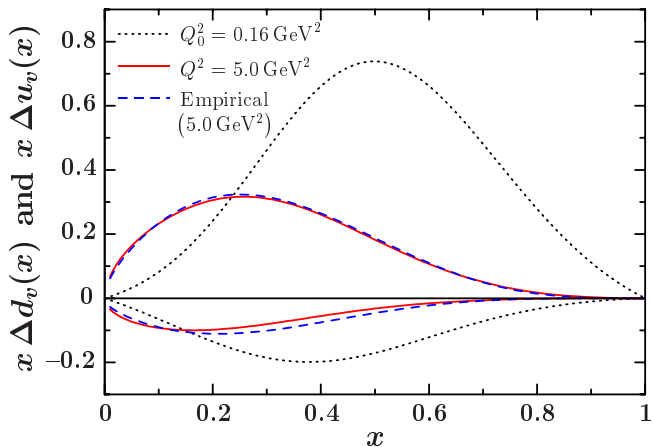}
\caption{Spin-dependent valence $u$ and $d$ distributions multiplied by Bjorken $x$.
The curves are as in Fig.~\ref{fig:SI_xup}, with empirical parameterizations
taken from Ref.~\cite{Hirai:2003pm}.}
\label{fig:SD_xup}
\end{figure}

Results for the spin-independent and spin-dependent 
valence $u$ and $d$ distributions
are presented in Figs.~\ref{fig:SI_xup} and \ref{fig:SD_xup}, respectively.
We show the predictions at the model scale 
and after QCD evolution\footnote{We utilize the computer program
of Ref.~\cite{Miyama:1995bd} for the spin-independent
case and of Ref.~\cite{Hirai:1997gb} for the spin-dependent case. 
We choose DGLAP evolution with
$N_f=3$, $\Lambda_{\text{QCD}}=250~$MeV in 
the $\overline{\text{MS}}$ renormalization scheme
up to NLO.} to $Q^2 = 5\,$GeV$^2$, where they are compared to empirical 
parameterizations.
We find excellent agreement between the model results and the parameterizations.
In comparison with the pure scalar model \cite{Mineo:2003vc,Cloet:2005tq}, 
the agreement has improved substantially, 
especially for the spin-dependent case. 

Our model results for the first polarized moments are 
$\D \, u_v = 0.924$ and $\D \, d_v = -0.343$
which agree with the values $\D \, u_v = 0.926 \pm 0.014$ and $\D \, d_v = -0.341 \pm 0.018$
determined from the axial coupling constants of octet baryons discussed in Ref.~\cite{Goto:1999by}. 
This emphasizes the importance of including axial-vector diquark correlations, since the pure scalar model 
would give a vanishing $\D \,d_v$ and a somewhat 
smaller $\D \,u_v$. The spin sum in our model
is $\D \S = 0.581$, which is smaller than the 
result of the pure scalar model, but
still somewhat larger than the accepted value of $\D \S = 0.213 \pm 0.138$ \cite{Hirai:2003pm}. This discrepancy
primarily reflects the absence of the $U$(1) axial anomaly \cite{Altarelli:1988nr,Efremov:1988zh}.

The behaviour of structure function and hence quark 
distribution ratios at large $x$ has been an area
of considerable debate \cite{Melnitchouk:1995fc,Zheng:2004ce} 
and is one of the regions where 
perturbative QCD (pQCD) offers firm predictions~\cite{Farrar:1975yb}.  
Experimentally, the ratio $d(x)/u(x)$ is surprisingly 
poorly known~\cite{Botje:1999dj}. 
In the limit $x \to 1$ it 
is thought to lie somewhere between $0$, the prediction based on scalar
diquark dominance \cite{Close:1973xw}  and $\tfrac{1}{5}$, the
pQCD result \cite{Farrar:1975yb}.
Analysis in Ref.~\cite{Melnitchouk:1995fc} favours the pQCD prediction. 
The same predictions also hold for the spin-dependent ratio, 
$\Delta d(x)/\Delta u(x)$, as $x$ approaches 1. 

\begin{figure}[tbp]
\centering\includegraphics[width=\columnwidth,angle=0]{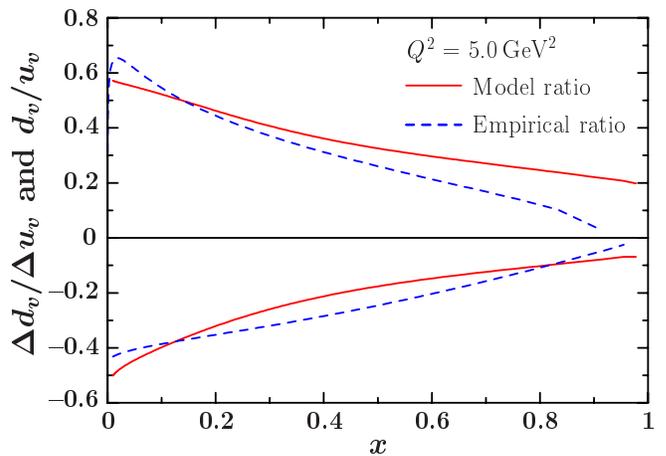}
\caption{Mixed flavour ratios for spin-independent and spin-dependent distributions.
There are two curves for each ratio, with the lower curves the polarized result.
The curves are as in Fig.~\ref{fig:SI_xup}, 
with spin-independent parameterizations
taken from Ref.~\cite{Martin:2002dr} and the 
spin-dependent from Ref.~\cite{Hirai:2003pm}}
\label{fig:SD_SI_du}
\end{figure}

In Fig.~\ref{fig:SD_SI_du} we plot our results for the ratios 
$d_v(x)/u_v(x)$ and $\D d_v(x)/\D u_v(x)$, together 
with the ratios of the empirical distributions. 
The $x \to 1$ limit of the spin-independent ratio 
is in agreement with the pQCD result.
The spin-dependent ratio, however, approaches $\sim -\tfrac{1}{16}$, 
the opposite sign to the pQCD prediction. Although the  
empirical parameterizations are constrained to 
give $0$ for these ratios as $x \to 1$,
we note that the systematic errors in both empirical 
ratios are very large in the region $x \gtrsim 0.5$
\cite{Blumlein:2002be,Martin:2002aw,Martin:2003sk,Hirai:2003pm}.
\begin{figure}[tbp]
\centering\includegraphics[width=\columnwidth,angle=0]{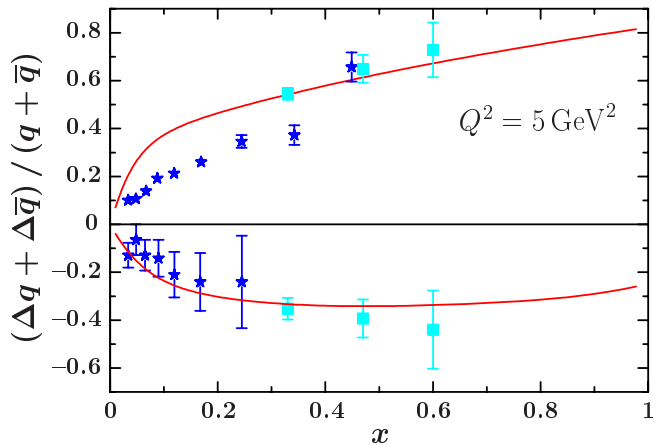}
\caption{Single flavour ratios $\lf(\D q + \D \ol{q}\rg)/\lf(q + \ol{q}\rg)$
where $q \in \lf(u,d\rg)$, at the scale $Q^2 = 5.0\,\text{GeV}^2$. The lower
curve represents the $d$-quark distribution.
The experimental results are from Hall A at Jefferson Lab \cite{Zheng:2004ce} (solid squares) and
Hermes \cite{Ackerstaff:1999ey} (solid stars).}
\label{fig:SD_SI_du_total}
\end{figure}

It is important to note that the pQCD predictions for 
the mixed flavour ratios are somewhat
model dependent, as assumptions have to be made about the relative 
strengths of the $u$ and $d$ contributions to the nucleon 
wavefunction. A more rigorous pQCD prediction,
relying only on helicity conservation, 
is possible for the single flavour ratios $\D u(x)/u(x)$ and $\D d(x)/d(x)$.
Perturbative QCD predicts that both these ratios 
should approach 1 for large $x$, which would require  
a change of sign in the $\D d$ distribution.

In Fig.~\ref{fig:SD_SI_du_total} we plot our results for the ratios 
$\lf(\D q + \D \ol{q}\rg)/\lf(q + \ol{q}\rg)$ where $q \in \lf(u,d\rg)$.
Since we wish to compare these ratios directly to 
recent experimental data, we include
sea quark distributions generated through the $Q^2$ evolution.  
In the $x \to 1$ limit our model
ratios approach  $\approx 0.8$ for the $u$-quark 
and $\approx -0.25$ for the $d$-quark. This seeming contradiction
to pQCD has also been suggested by recent experiments 
by the Jefferson Lab Hall A collaboration
\cite{Zheng:2004ce,Zheng:2003un},
with our predictions consistent with their experimental 
results. This data is also shown in Fig.~\ref{fig:SD_SI_du_total}.
\begin{figure}[tbp]
\centering\includegraphics[width=\columnwidth,angle=0]{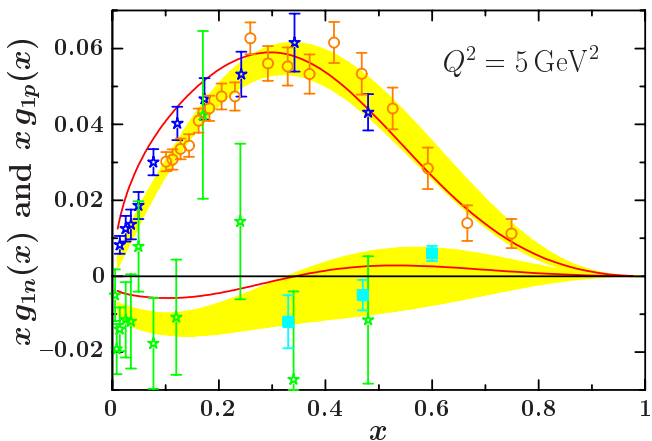}
\caption{Polarized structure functions $g_{1p}$ and $g_{1n}$ at $Q^2 = 5\,$GeV$^
2$.
The solid line is the model prediction, with
the lower curve corresponding to $g_{1n}$. The shaded areas
represent the empirical parameterizations with uncertainties of
Ref.~\cite{Blumlein:2002be}, at the same scale. The experimental data, with
$1 \leq Q^2 \leq 10\,$GeV$^2$, is from
SMC \cite{Adeva:1998vw} (open stars),
SLAC E143 \cite{Abe:1998wq} (open circles) and JLab \cite{Zheng:2004ce} (solid squares).}
\label{fig:g1pg1n}
\end{figure}

In Fig.~\ref{fig:g1pg1n} we give our results for the 
spin-dependent structure functions
$g_{1p}(x)$ and $g_{1n}(x)$. 
The parameterizations of Ref.~\cite{Blumlein:2002be} 
are also included as the shaded areas,
which indicate the empirical uncertainties. 
Our results compare well with the empirical parameterizations, 
agreeing within uncertainties for the 
region $x \gtrsim 0.25$. Comparison with experiment is also favorable,
although the experimental determination for $g_{1n}(x)$ is less certain. 

\begin{figure}[tbp]
\centering\includegraphics[width=\columnwidth,angle=0]{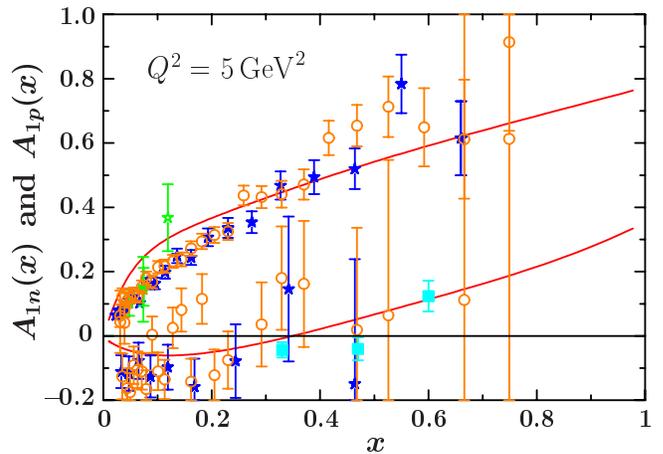}
\caption{Structure function ratios $A_{1p}$ and $A_{1n}$, at the scale $Q^2 = 5\
,$GeV$^2$.
The solid line is the model prediction, with the lower line corresponding to $A_
{1n}$.
The experimental points represent data with
$1 \leq Q^2 \leq 10\,$GeV$^2$ from Hermes \cite{Airapetian:1998wi} (closed star)
, SMC  \cite{Adeva:1998vw}
(open star), E143 \cite{Abe:1998wq} (open circles) and JLab \cite{Zheng:2004ce}
(solid squares).
We do not plot the E155 \cite{Anthony:2000fn} results as they are similar to those from E143.}
\label{fig:A1pA1n}
\end{figure}

Model results for the asymmetries $A_{1p}(x)$ and $A_{1n}(x)$ 
are given in Fig.~\ref{fig:A1pA1n}, where we see 
good agreement in the valence region. However, for   
small $x$, $A_{1p}(x)$ is slightly too large, because 
of the enhancement of $g_{1p}(x)$
in the same region. This is most likely associated with the omission of 
the effects of the axial anomaly in the present work. 
It is also clear from the experimental 
data that the uncertainties in these
ratios at large $x$, are still significant.

\vspace{-0.5em}
\section{Conclusion}
\vspace{-0.2em}

Using a covariant quark-diquark model for the nucleon,
including both scalar and axial-vector diquark channels, we calculated
the spin-independent and spin-dependent quark light-cone momentum distributions 
and structure functions.
A key feature of the framework is that it produces quark distributions 
that have the correct support and obey the number and momentum sum rules.
The model also incorporates important aspects of confinement by
eliminating unphysical thresholds for nucleon decay into quarks.

Highlights of our results are obtaining values for the polarized first moments
of the quark distributions $\D u_v = 0.924$ and $\D d_v = -0.343$, in agreement with 
those obtained from axial couplings of octet baryons. 
We  also obtain excellent agreement 
with empirical parameterizations of the valence quark distributions.
We paid special attention to the single flavour ratios 
$\lf(\D q + \D \ol{q}\rg)/\lf(q + \ol{q}\rg)$ and the asymmetries
$A_{1p}$ and $A_{1n}$, finding good agreement with 
recent experimental results from JLab. 

These results indicate that diquark correlations are an essential feature of the 
non-perturbative structure of the nucleon. In particular,
the admixture of axial-vector diquarks,
though small, is essential to obtain the observed agreement
with empirical data.  

Finally, we would like to mention that a very 
important advantage of this
covariant quark-diquark model is that it can be 
readily extended to the case of finite
nucleon density. The results presented in this paper 
strongly suggest that this model should provide a reliable basis from which 
to begin investigation of the medium modifications of 
both spin-independent and spin-dependent structure functions.

\vspace{-1.5em}
\section*{Acknowledgments} 
\vspace{-0.2em}

IC and WB thank W. Melnitchouk and H. Mineo for interesting and helpful discussions.
This work was supported by the Australian Research Council and DOE 
contract DE-AC05-84ER40150,
under which SURA operates Jefferson Lab, and by the Grant in Aid for Scientific
Research of the Japanese Ministry of Education, Culture, Sports, Science and
Technology, Project No. C2-16540267.

\end{document}